\documentclass{ieeeaccess}
\usepackage{cite}
\usepackage{amsmath,amssymb,amsfonts}
\usepackage[nolist]{acronym}
\usepackage{algorithmic}
\usepackage{booktabs}
\usepackage{graphicx}
\usepackage{textcomp}
\usepackage{xfrac}
\usepackage{wrapfig}

\usepackage[normalem]{ulem}

\usepackage{bm}
\makeatletter
\AtBeginDocument{\DeclareMathVersion{bold}
\SetSymbolFont{operators}{bold}{T1}{times}{b}{n}
\SetSymbolFont{NewLetters}{bold}{T1}{times}{b}{it}
\SetMathAlphabet{\mathrm}{bold}{T1}{times}{b}{n}
\SetMathAlphabet{\mathit}{bold}{T1}{times}{b}{it}
\SetMathAlphabet{\mathbf}{bold}{T1}{times}{b}{n}
\SetMathAlphabet{\mathtt}{bold}{OT1}{pcr}{b}{n}
\SetSymbolFont{symbols}{bold}{OMS}{cmsy}{b}{n}
\renewcommand\boldmath{\@nomath\boldmath\mathversion{bold}}}
\makeatother

\def\BibTeX{{\rm B\kern-.05em{\sc i\kern-.025em b}\kern-.08em
    T\kern-.1667em\lower.7ex\hbox{E}\kern-.125emX}}

\newcommand{\commenttxt}[1]{{#1}}

\begin{document}
\begin{acronym}
        \acro{3GPP}{3rd generation partnership project}
        \acro{5G}{fifth generation of mobile networks}
        \acro{6G}{sixth generation of mobile networks}
        \acro{AO}{alternating optimization}
	\acro{AP}{access point}
        \acro{AWGN}{additive white Gaussian noise}
        \acro{DFT}{discrete Fourier transform}
        \acro{ETSI}{European Telecommunications Standards Institute}
        \acro{HWI}{hardware imperfection}
        \acro{LOS}{line-of-sight}
        \acro{MIMO}{multiple-input multiple-output}
        \acro{MISO}{multiple-input single-output}
        \acro{MLP}{multilayer perceptron}
        \acro{MSE}{mean squared error}
        \acro{NLOS}{non-line-of-sight}
        \acro{NN}{neural network}
        \acro{OFDM}{orthogonal frequency division multiplex}
        \acro{OTA}{over-the-air}
        \acro{PSN}{phase-shift noise}
        \acro{ReLU}{rectified linear unit}
        \acro{REV}{Rotating Element Vector}
        \acro{RIS}{reconfigurable intelligent surface}
        \acro{SCA}{successive convex approximation}
        \acro{STM}{strongest tap maximization}
        \acro{UE}{user equipment}
        \acro{VNA}{vector network analyzer}
\end{acronym}
\history{Date of publication xxxx 00, 0000, date of current version xxxx 00, 0000.}
\doi{10.1109/ACCESS.2024.0429000}

\title{Lightweight Gradient Descent Optimization for Mitigating Hardware Imperfections in RIS Systems}
\author{\uppercase{Pedro H. C. de Souza}\authorrefmark{1},
\uppercase{Luiz A. M. Pereira}\authorrefmark{1}, \uppercase{Faustino R. Gómez}\authorrefmark{1}, \uppercase{Elsa M. Materón}\authorrefmark{1}, \uppercase{Jorge Ricardo Mejía-Salazar}\authorrefmark{1}, and \uppercase{Luciano Mendes}\authorrefmark{1}}

\address[1]{National Institute of Telecommunications - Inatel, Santa Rita do Sapucaí, MG 37536-001 Brazil}
\tfootnote{This work has been funded by the following research projects: Brasil 6G Project with support from RNP/MCTI (Grant 01245.010604/2020-14), xGMobile Project code XGM-AFCCT-2024-2-15-1 with resources from EMBRAPII/MCTI (Grant 052/2023 PPI IoT/Manufatura 4.0) and FAPEMIG (Grant PPE-00124-23), SEMEAR Project supported by FAPESP (Grant No. 22/09319-9), SAMURAI Project supported by FAPESP (Grant 20/05127-2), Ciência por Elas with resources from FAPEMIG (Grant APQ-04523-23), Fomento à Internacionalização das ICTMGs with resources from FAPEMIG (Grant APQ-05305-23), Programa de Apoio a Instalações Multiusuários with resources from FAPEMIG (Grant APQ-01558-24), and Redes Estruturantes, de Pesquisa Científica ou de Desenvolvimento Tecnológico with resources from FAPEMIG (Grant RED-00194-23). This work has also been supported by a fellowship from CNPq and FAPESP.}

\markboth
{Author \headeretal: Preparation of Papers for IEEE TRANSACTIONS and JOURNALS}
{Author \headeretal: Preparation of Papers for IEEE TRANSACTIONS and JOURNALS}

\corresp{Corresponding author: Pedro H. C. de Souza (e-mail: pedro.carneiro@dtel.inatel.br).}

\begin{abstract}
Ongoing discussions about the future of wireless communications are reaching a turning point as standardization activities for the \ac{6G} become more mature. New technologies must now face renewed scrutiny by the industry and academia in order to be ready for deployment in the near future. Recently, \acp{RIS} gained attention as a promising solution for improving the propagation conditions of signal transmission in general. The \ac{RIS} is a planar array of tunable resonant elements designed to dynamically and precisely manipulate the reflection of incident electromagnetic waves. However, the physical structure of the \ac{RIS} and its components may be subject to practical limitations and imperfections. It is imperative that the \acp{HWI} associated with the \ac{RIS} be analyzed, so that it remains a feasible technology from a practical standpoint. Moreover, solutions for mitigating the \acp{HWI} must be considered, as is discussed in this work. More specifically, we introduce a gradient descent optimization for mitigating \acp{HWI} in \ac{RIS}-aided wideband communication systems. Numerical results show that the proposed optimization is able to compensate for \acp{HWI} such as the \ac{PSN} and \ac{RIS} surface deformations.
\end{abstract}


\titlepgskip=-21pt

\maketitle

\section{Introduction}
\acresetall
\label{sec:introduction}
\PARstart{W}{ith} the consolidation of the \ac{5G} and the ongoing standardization efforts in the \ac{6G}, emerging technologies such as \acp{RIS} have become the focus of studies in industry and academia alike \cite{astr:24,alexandropoulos:22}. The \ac{RIS} is typically envisioned as a planar surface, built with passive elements, such as tunable resonant elements with adjustable impedance values (phase-shifts) \cite{liaskos:22}. This enables the \ac{RIS} elements or reflectors, to act as independent wave scatterers \cite{bjorn:22}. In other words, a well positioned \ac{RIS} between a transmitter and receiver can create additional paths for the signal propagation. Consequently, favorable propagation conditions for the signal transmission can be established even in scenarios where the transmitter have a \ac{NLOS} in relation to the receiver.

\commenttxt{Although numerous works in the literature address \ac{RIS} systems, the majority focus on idealized system-level operations and performance metrics, such as maximizing the channel capacity at the receiver \cite{yang:20}. In contrast, the authors of \cite{wang:24} investigate a specific \ac{RIS} hardware implementation in detail, describing the design of individual reflector elements and analyzing their wave propagation characteristics. However, studies that jointly consider both system-level performance and the practical challenges introduced by \acp{HWI} remain relatively scarce. For instance, \cite{yang:23} experimentally investigates the beamforming behavior of a \ac{RIS} and subsequently develops an analytical model to assess the impact of different \acp{HWI} on system performance. Alternatively, \cite{yue:24} accounts for noise effects in the configuration of \ac{RIS} elements. The relevance of addressing \acp{HWI} in the context of \ac{RIS} systems is further emphasized in \cite{xian:25}, which argues that they are inevitable in practice. Recently, instead of merely analyzing the performance degradation caused by \acp{HWI}, the authors of~\cite{huang:24} proposed a novel gradient descent–based optimization framework to actively compensate for phase-shift errors and phase-dependent amplitude coupling in individual \ac{RIS} reflector elements. Representative examples of \acp{HWI} include, but are not limited to, amplitude–phase correlation, quantization errors, phase noise, amplifier non-linearity, carrier frequency offset, and unavoidable manufacturing deviations. Consequently, effective calibration and compensation of the aforementioned \acp{HWI} are considered essential for the practical deployment of future \ac{RIS} systems. As reported in \cite{liX:25}, the implementation and practical considerations of \ac{RIS} systems are already recognized as a key item in the ongoing \ac{ETSI} standardization efforts.}



In this work, we build upon the approach presented in~\cite{huang:24} to propose a method for mitigating the impact of \acp{HWI} on \ac{RIS} performance. However, here we leverage numerical tools and automatic differentiation as a means to compute the gradient descent. This contrasts to the approach taken by the authors of \cite{huang:24}, where the expression for calculating the gradients is derived analytically, requiring the knowledge a priori of the distributions that describe the behavior of the \acp{HWI}. Furthermore, the method of \cite{huang:24} also requires a numerical computation of integrals, thus a closed-form solution is at any rate not provided. \commenttxt{These requirements increase computational complexity and limit adaptability to dynamic channel conditions, posing significant challenges for real-time or large-scale \ac{RIS} deployment. Therefore, the main objective of our work lies in proposing a gradient descent optimizer for mitigating \acp{HWI} in \ac{RIS} systems, albeit one that exclusively relies on instantaneous parameters, such as the channel coefficients.}

\commenttxt{More specifically, this work introduces a numerically driven gradient descent optimization framework for mitigating \acp{HWI} in \ac{RIS} assisted systems. The proposed method leverages automatic differentiation to compute gradients directly from instantaneous system parameters, eliminating the need for analytical derivations or prior statistical modeling of \acp{HWI}. We consider the effects of the \ac{PSN} and \ac{RIS} surface deformations in the context of signal transmission, and reflection, over wideband communication channels, by using the \ac{OFDM} system. To the best of authors' knowledge, this is the first work that jointly considers the \ac{PSN} and \ac{RIS} surface deformations in the context of \ac{RIS} systems with wideband signal reflection. Moreover, we also provide complementary experimental evaluations of the \ac{RIS} electromagnetic response under surface deformations, in order to underscore the importance of taking into account practical \acp{HWI}.} In summary, the main contributions of this work can be established as follows:
\begin{itemize}
   \item \ac{RIS} phase-shift compensation with no prior knowledge of \acp{HWI} distributions;
   \item \commenttxt{Costly computations of expectation values with respect to the \acp{HWI} distributions \cite{yue:24,huang:24} are not required;}
   \item The \commenttxt{joint} effect of surface deformations \commenttxt{and \ac{PSN}} on the \ac{RIS} performance are investigated;
   \item The gradient descent is computed considering multiple \ac{OFDM} subcarriers. 
\end{itemize}

The manuscript is organized as follows: Section \ref{sec:sysmodel} describes the channel model and metrics used in the context of the discussed \ac{RIS}-aided system; Section \ref{sec:hdwImperf} details the \acp{HWI} modeling; Section \ref{sec:grad} specifies the assumptions for the gradient coefficients computations; Section \ref{sec:expresult} discusses the results obtained from a practical experiment involving the \ac{RIS} prototype; Section \ref{sec:numresult} delves into the principles behind the proposed gradient descent optimization and the \ac{RIS} performance results under compensated \acp{HWI} are also analyzed, and finally, Section \ref{sec:conclusion} concludes the paper.


\subsection{Notation}
Throughout this work, italicized letters (e.g. $x$ or $X$) represent scalars, boldfaced lowercase letters (e.g. $\mathbf{x}$) represent vectors, and boldfaced uppercase letters (e.g. $\mathbf{X}$) denote matrices. The $n$th entry of the vector $\mathbf{x}$ is represented by $x\left[n\right]$. The entry on the $i$th row and $j$th column of the matrix $\mathbf{X}$ is denoted by $X_{i,j}$. The sets of real and complex numbers are represented by $\mathbb{R}$ and $\mathbb{C}$, respectively. The phase content of $x \in \mathbb{C}$ is given by $\arg\left\{x\right\}$ in radians. The $\lfloor x\rceil$ operator denotes the rounding of scalars to the nearest integer. The sets of vectors of dimension $X$ with real and complex entries are respectively represented by $\mathbb{R}^{X}$ and $\mathbb{C}^{X}$. The sets of matrices of dimension $X\times Y$ with real and complex entries are correspondingly described by $\mathbb{R}^{X\times Y}$ and $\mathbb{C}^{X\times Y}$. The transposition and conjugate transposition operations of a vector or matrix are represented as $\left(\cdot\right)^{\text{T}}$ and $\left(\cdot\right)^{\text{H}}$, respectively. The $\ell_p$-norm, $p \geq 1$, of the vector $\mathbf{x}$ is given by $\|\mathbf{x}\|_p = \left(\lvert x\left[0\right]\rvert^p + \lvert x\left[1\right]\rvert^p + \cdots + \lvert x\left[n - 1\right]\rvert^p\right)^{1/p}$. We also have $\left\langle\mathbf{x}\right\rangle = X^{-1}\sum_n x\left[n\right]$ for $\mathbf{x} \in \mathbb{R}^{X}$. 

\section{System Model}\label{sec:sysmodel}
Consider a metasurface composed by passive components that can reflect the impinging electromagnetic waves. These components can be thought as varactors or varistors with adjustable impedance values, so that each component constitutes an element or reflector of a \ac{RIS}. Typically the \ac{RIS} reflectors are arranged in a planar of \commenttxt{$N$} reflectors (with sides of size $d_\text{H}$ and $d_\text{V}$ meters each), which can be represented mathematically by $\boldsymbol{\omega}_{\theta} = e^{\jmath \boldsymbol{\theta}} \in \mathbb{C}^{N}$, where $\boldsymbol{\theta} = [\theta_0\text{,}\theta_1\text{, }\dots\text{,}\theta_{N - 1}]^\text{T}$. \commenttxt{Note that $\boldsymbol{\omega}_{\theta}[n] = e^{\jmath \theta_n}$ gives the complex number representing the configuration of the $n$-th \ac{RIS} reflector, for which the phase-shift is expressed by $\theta_n = \left[-\pi,\pi\right] \ \forall n \in N$.} Therefore, a receiving \ac{UE} can experience improvements in communication, whenever the signal transmitted by the \ac{AP} is properly reflected by the \ac{RIS}.

\begin{figure}[h!]
	\centering
	\includegraphics[width=0.775\linewidth,keepaspectratio]{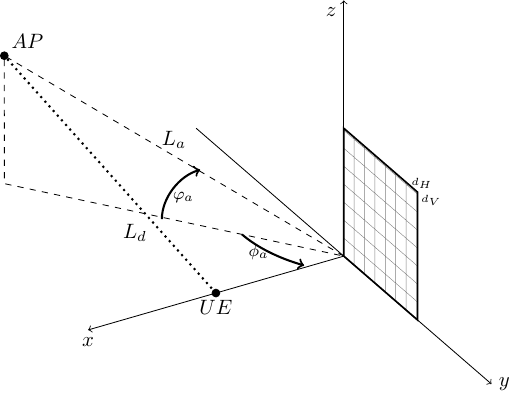}
	\caption{Spatial diagram of the propagation model on a $\left(x\text{, } y\text{, } z\right)$ coordinate system. Each \ac{RIS} reflector has sides of size $d_\text{H}$ and $d_\text{V}$ meters, as illustrated.}\label{fig:sysdiag}
\end{figure}

To elaborate, assume the baseline propagation model of Figure~\ref{fig:sysdiag}. It shows the direct channel with $L_d$ propagation paths between the \ac{AP} and \ac{UE}; the composite channel is also illustrated, being composed by the channels cascade between the \ac{AP} and \ac{RIS}, with $L_a$ paths, and \ac{RIS} to the \ac{UE}, with $L_b$ paths (omitted from Figure~\ref{fig:sysdiag} for better overall visibility). The direct channel is assumed to be \ac{NLOS} whereas the composite channel is \ac{LOS} dominated. This propagation scenario may occur when the \ac{UE} is experiencing signal blocking, while the \ac{RIS} is not, since it can be placed in an advantageous position for signal reflection \cite{bjorn:22,liaskos:22}. In summary, the main objective of the \ac{RIS} is to change phase rotations of the signal propagating through the composite channel, in order to combine coherently at the \ac{UE} the signals coming from both the direct and composite channels. More specifically, for each different path of the composite channel ($L_a L_b$ paths), $N$ new paths are created with a modified phase-shift.

Also illustrated in Figure~\ref{fig:sysdiag} are the relative azimuth angles of arrival ($\phi_a$) and departure ($\phi_b$) at/from the \ac{RIS}, as well as the respective elevation angles $\varphi_a$ and $\varphi_b$ (angles for the $L_b$ channel paths are omitted given the redundant representations). These angles are used to compute the array response of the \ac{RIS} reflectors as a whole. The array response dictates that the \ac{RIS} reflector farthest from the plane origin imposes more rotation to the signal, with the other reflectors presenting gradually less rotation as they draw closer to the plane origin \cite{pedro:25}. These rotation values are then weighted by the impinging signal angles, to compensate for different signal source/destination locations on the plane of Figure~\ref{fig:sysdiag}. \commenttxt{Furthermore, note that we assume only the propagation of plane electromagnetic waves, corresponding to the far-field region, where such waves arise from constructive and destructive phase-interference among the radiated fields. This assumption inherently accounts for the averaged response of the \ac{RIS} without explicitly modeling near-field coupling between elements, as shown by \cite{jorge:25}.}

In this work we also consider that signal transmission is performed over wideband communication channels, by resorting to the \ac{OFDM} system. This adds to the difficulty of finding the appropriate configuration for the \ac{RIS} reflectors. It is known that there is no unique phase-shift configuration for the \ac{RIS} that can simultaneously maximize the channel capacity at the \ac{UE} for all $K$ subcarriers \cite{bjorn:22}. Bear in mind that each subcarrier presents different channel phase rotations, rendering it unfeasible to configure a different $\boldsymbol{\omega}_{\theta}$ for each subcarrier whilst a constant \ac{RIS} configuration is assumed for a given bandwidth. Therefore, to further understand the \ac{RIS} configuration problem, we define the achievable rate \cite{bjorn:22,pedro:25} as follows
\begin{equation}\label{eq:rate}
    R = \frac{B}{\xi}\sum_{i=0}^{K - 1}{\log_2{\left(1 + \frac{p_i\|\mathbf{f}_i^\text{H} \mathbf{h}_d + \mathbf{f}_i^\text{H} \mathbf{V}^{\text{T}}\boldsymbol{\omega}_{\theta}\|_2^2}{BN_0}\right)}} \ \sfrac{\text{bit}}{\text{s}}\text{,}
\end{equation}
wherein $\xi = K + M - 1$, to take into account the cyclic prefix loss, $\mathbf{p} \in \mathbb{R}^{K}$ is the power vector, with $p\left[k\right]$ being the power allocated to the $k$-th subcarrier, such that $P = \left\langle \mathbf{p}\right\rangle$; $P$ being the total transmission power, $\mathbf{f}_i$ represents the $i$-th row of the \ac{DFT} matrix $F_{i,j} = e^{-\jmath 2\pi ij/K}$, $\mathbf{h}_d \in \mathbb{C}^{K}$ denotes $K$ samples (with $K - M$ padding samples) of the discrete-time impulse response of the direct channel, $\mathbf{V} \in \mathbb{C}^{N \times K}$ describes the discrete-time impulse response for all $N$ composite channels, $B$ is the total bandwidth occupied by $K$ subcarriers, and finally, $N_0$ is the \ac{AWGN} power density.

Notice in \eqref{eq:rate} that the \ac{RIS} configuration has considerable influence on the degree to which a coherent or constructive combination of signals can be achieved at the \ac{UE}. By letting the subcarriers power allocation be provided by the well-known water filling algorithm \cite{bjorn:13,yang:20}, then it is easy to show that a more constructive combination of signals is directly proportional to higher values for the achievable rate of \eqref{eq:rate}. In fact, a wealth of research \cite{bjorn:22,feng:21,pedro:25} is available in which the optimization of the \ac{RIS} configuration is investigated in the context of achievable rate maximization, for instance. However, this matter is not as straightforward when hardware impairments or imperfections prevent the \ac{RIS} to function within its nominal parameters. In the next section we delve into more details about \ac{RIS} hardware imperfections and propose a solution to mitigate their undesirable effects.

\section{RIS Hardware Imperfections}\label{sec:hdwImperf}
The \ac{RIS} configuration may deviate from the ideal due to a variety of hardware imperfections or impairments \cite{yang:23}. Typically, the \ac{RIS} can be contaminated mainly by system-level noise as well as hindered by imperfections on the material that constitutes its surface, for example. Therefore, we first present the \ac{PSN} model, followed by the modeling of \ac{RIS} surface deformations and their consequences on the \ac{RIS} configuration. 

\subsection{Phase-shift Noise}
In this work, we employ the following modeling for the \ac{PSN} impairment:
\begin{equation}\label{eq:errmodel}
    \boldsymbol{\hat{\omega}}_{\theta} = \epsilon \boldsymbol{\omega}_{\theta} + \boldsymbol{\upsilon}\sqrt{1 - \epsilon^2};
\end{equation}
for which $\boldsymbol{\hat{\omega}}_{\theta} \in \mathbb{C}^N$ represents an imperfect configuration for the \ac{RIS} phase-shifts due to the \ac{PSN}: ${\boldsymbol{\upsilon} = [\upsilon_0\text{,}\upsilon_1\text{, }\dots\text{,}\upsilon_{N - 1}]^\text{T} \in \mathbb{R}^N}$, wherein each entry is given by $\upsilon_n \sim \mathcal{N}\left(0,1\right), \forall \ n$. More specifically, note in \eqref{eq:errmodel} that the uncorrelated \ac{PSN} values, $\boldsymbol{\upsilon}$, are being combined with the ideal configured phase-shifts, $\boldsymbol{\omega}_{\theta}$, accordingly, so that $\|\hat{\omega}_{\theta_n}\|_2^2 = 1 \ \forall  n$. In other words, the inherent passivity of the \ac{RIS} reflectors remains preserved. Therefore, a wide range of \ac{PSN} intensity can be investigated by adjusting $\epsilon \in \mathbb{R}$, since $\epsilon = 0$ represents the worst configuration for the \ac{RIS}, whereas $\epsilon = 1$ is the ideal case where there are no imperfections.

\subsection{RIS Surface Deformations}
\begin{figure}[h!]
	\centering
	\includegraphics[width=0.45\linewidth,keepaspectratio]{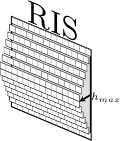}
	\caption{\ac{RIS} surface deformation (not to scale) considering $k = \pi$ in \eqref{eq:defmodel}. Notice how grouped fixed deformations occur for all reflectors of a same (horizontal) line.}\label{fig:defdiag}
\end{figure}

Whether the \ac{RIS} planar surface is subjected to deterioration due to the environment exposure or even because of imprecisions in the manufacturing process, it can nevertheless cause a considerable negative impact on the \ac{RIS} performance \cite{yang:23}. In this work, we consider \ac{RIS} surface deformations of a regular shape, giving rise to fixed phase-shift errors for groups of reflectors. The grouped fixed phase-shift errors are given by
\begin{align}\label{eq:defmodel}
    \boldsymbol{\hat{\theta}} &= \boldsymbol{\theta} + \frac{2\pi}{\lambda} h_{max}\sin{\left(k\boldsymbol{\psi}\right)}\left(\cos{\left(\varphi_a\right)} + \cos{\left(\varphi_b\right)}\right)\text{; where} \\
    \boldsymbol{\psi} &= \left(N_\text{col} - 1\right)^{-1}\left[\lfloor\sfrac{0}{N_\text{col}}\rceil \text{,}\lfloor\sfrac{1}{N_\text{col}}\rceil \text{,}\dots\text{,}\lfloor\sfrac{N-1}{N_\text{col}}\rceil\right]^\text{T}\text{,} \nonumber
\end{align}
in which $\boldsymbol{\theta} \in \mathbb{R}^N$ is the ideal phase-shift configuration, also $\lambda = \commenttxt{c} f_c^{-1}$ \commenttxt{($c = 3 \times 10^8$ (m/s))}, $f_c$ being the central frequency of the signal carrier, $h_{max}$ represents the maximum deformation or displacement of the \ac{RIS} surface in relation to the ideal surface and $k$ defines the number of peaks (maximum) deformations across the \ac{RIS} surface. Consequently, $\boldsymbol{\omega}_{\hat{\theta}}$ represents the \ac{RIS} configuration affected by the surface deformations. Figure~\ref{fig:defdiag} illustrates an example of the surface deformations that result from $k = \pi$ in \eqref{eq:defmodel}.

\section{Gradient Computations}\label{sec:grad}
Suppose that the \ac{RIS} is able to afford local signal processing\commenttxt{\footnote{\commenttxt{For simplification, the optimization processing is not offloaded to the \ac{AP}.}}}, such as the following gradient computations
\begin{equation}\label{eq:grad}
    \nabla_{\boldsymbol{\bar{\theta}}} =-\sum_{\forall i}\frac{\partial \|\mathbf{f}_i^\text{H} \mathbf{h}_d + \mathbf{f}_i^\text{H} \mathbf{V}^{\text{T}}\boldsymbol{\omega}^\prime_{\theta}\|_2^2}{\partial \boldsymbol{\omega}^\prime_{\theta}}\Biggr\rvert_{\boldsymbol{\omega}^\prime_{\theta} = \boldsymbol{\bar{\omega}}_\theta} \text{,}
\end{equation}
in which $\boldsymbol{\omega}^\prime_{\theta} = e^{\jmath \boldsymbol{\theta}^\prime} \in \mathbb{C}^N$ represents the \ac{RIS} configuration affected by the hardware imperfections discussed in Section~\ref{sec:hdwImperf}, while $\boldsymbol{\bar{\omega}}_\theta = e^{\jmath(\boldsymbol{\theta}^\prime + \boldsymbol{\bar{\theta}})} \in \mathbb{C}^N$ is denoted as the compensated \ac{RIS} configuration. Therefore, we propose a lightweight gradient descent optimization based on the gradients computed via \eqref{eq:grad}. It can be used to compute phase-shifts compensations ($\boldsymbol{\bar{\theta}} \in \mathbb{R}^N$), so that the \ac{RIS} hardware imperfections are locally compensated. 

Notice in \eqref{eq:grad} that the \ac{RIS} is assumed to be channel-aware, since the gradient computations make use of channel samples $\mathbf{h}_d$ and $\mathbf{V}$. In addition, the imperfect configuration, $\boldsymbol{\omega}^\prime_{\theta}$, is also assumed to be known through the realization of \ac{RIS} calibration procedures. \commenttxt{One possible approach to realize the calibration is through \ac{OTA} calibration techniques, such as the method proposed in~\cite{zhang2024phases}. This approach establishes a solid theoretical foundation for estimating \ac{RIS} element phase deviations via a backpropagation-based optimization framework. Nevertheless, this method is computationally demanding and relies on extensive pilot-based external measurements. On the other hand, certain techniques originally developed for phased-array antenna calibration, such as the \ac{REV}, fast amplitude-only, and complex amplitude calibration methods \cite{pan2023phased}, could potentially be adapted and extended to \ac{RIS} systems, offering new directions for practical, scalable, and adaptive calibration procedures.}

\begin{figure}[h!]
	\centering
	\includegraphics[width=0.5\linewidth,keepaspectratio]{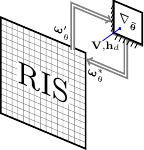}
	\caption{Diagram illustrating the compensation of \ac{RIS} hardware imperfections with the aid of gradient descent operations. The final adjusted \ac{RIS} configuration is denoted by $\boldsymbol{\omega}^*_{\theta}$.}\label{fig:gradDiag}
\end{figure}

Figure~\ref{fig:gradDiag} brings a diagram showing the compensation of the imperfect \ac{RIS} configuration from a general perspective.

The gradients computation, described symbolically in \eqref{eq:grad}, is carried out by numerical methods detailed in Section~\ref{sec:numresult}. It can be shown that the proposed gradient descent optimization is capable of minimizing the norm magnitude of \eqref{eq:grad}. Consequently, as discussed in Section~\ref{sec:sysmodel}, this is equivalent to maximizing the achievable rate at the \ac{UE} with the adjusted \ac{RIS} configuration $\boldsymbol{\omega}^*_{\theta}$.

\section{Experimental Results}\label{sec:expresult}
Prior to presenting the numerical results obtained with computational simulations, we discuss experimental insights on the \ac{RIS} electromagnetic response under surface imperfections. We briefly demonstrate in this section, that structural deformations, whether introduced during fabrication or installation, can result in undesirable frequency responses by the \ac{RIS}. The measurements were conducted in an anechoic chamber, ensuring that the \ac{RIS} was tested in a multipath-free environment. 

The \ac{RIS} prototype used in the experiment was developed with a PET (polyethylene terephthalate) substrate that was initially covered with adhesive paper as a masking layer. The desired metasurface pattern was then engraved onto the adhesive paper using a laser cutter. After the laser engraving, the excess adhesive material was carefully removed, leaving behind a patterned mask that exposed only the regions intended for metallization. Subsequently, silver conductive paint was applied to the exposed areas with a squeegee, using a screen-printing technique. The painted substrate was then cured in an oven at $40$ °C for $40$ minutes to ensure proper adhesion and conductivity. Once the curing process was complete and the paint had dried thoroughly, the remaining adhesive mask was peeled away, revealing the final metasurface structure defined by the silver coating. However, it is noteworthy that due to the high power of the laser cutter, the PET substrate exhibited several surface imperfections (primarily a high roughness) caused by heat exposure.  

\begin{figure}[h!]
    \centering
    \includegraphics[width=0.975\linewidth,keepaspectratio]{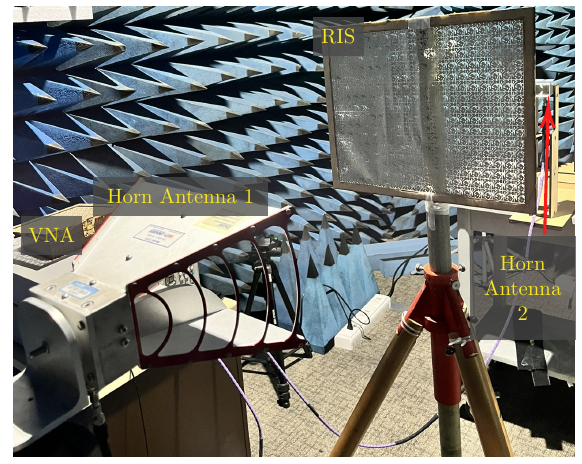}
    \caption{Photography of the experimental setup.}
    \label{fig:experimentalSetup}
\end{figure}

Figure~\ref{fig:experimentalSetup} illustrates the experimental setup, where we employed two horn antennas, the aforementioned \ac{RIS} prototype, and a Keysight E5071C \ac{VNA} to measure the transmission coefficient $S_{21}$. It is well known that the scattering parameter $S_{21}$ characterizes the forward transmission of a two-port \ac{VNA}, representing the proportion of the incident signal at port $1$ that is transmitted to port $2$ under matched termination conditions. This parameter encompasses both the amplitude and phase of the transmitted wave and is fundamental for assessing key performance metrics such as gain, insertion loss, and overall transmission efficiency. Typically, $|S_{21}|$ is expressed in decibels (dB) using a logarithmic scale. It is also widely accepted that when $S_{21}$ falls below $-10$ dB, the device under test exhibits predominantly reflective behavior, indicating poor transmission through the system~\cite{pozar2021microwave}. Moreover, note in Figure~\ref{fig:experimentalSetup} that the \ac{RIS} prototype was placed at the midpoint between the antennas, which were aligned to maintain consistent polarization and minimize misalignment losses. The ETS-Lindgren 3115 double-ridge horn antenna was used at the transmission side, while a custom-designed horn antenna was employed at the reception side.

\begin{figure}[h!]
    \centering
    \includegraphics[width=0.95\linewidth,keepaspectratio]{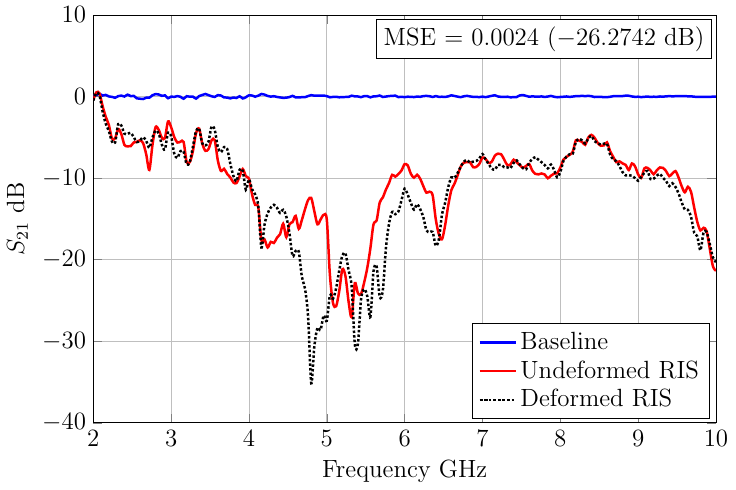}
    \caption{Measured transmission coefficient $S_{21}$ for the undeformed and deformed \ac{RIS}, highlighting the impact of structural modifications on the \ac{RIS}.}
    \label{fig:s21}
\end{figure}

The initial step in the measurement procedure consisted of a full two-port calibration of the \ac{VNA}, employing a standard calibration kit comprising of open, short, and matched load terminations. Following the calibration, the transmission coefficient $S_{21}$ was measured with the absence of the \ac{RIS} prototype between the horn antennas. This measurement was then stored in the \ac{VNA} internal memory, to be used subsequently as a baseline reference. Therefore, to isolate the frequency response of the \ac{RIS}, a normalization procedure was applied, in which each $S_{21}$ measurement was divided by the aforementioned baseline measurement. This operation is performed using the \ac{VNA} and effectively centers the resulting $S_{21}$ curves around $0$ dB, as shown in Figure~\ref{fig:s21}. Consequently, systematic measurement variations, path loss, and antennas response are compensated accordingly. Bear in mind that the $S_{21}$ measurement, after this calibration process, constitutes only of the measurement obtained from the \ac{RIS} prototype under evaluation.

Following the calibration process, the \ac{RIS} was positioned between the two horn antennas to acquire its $S_{21}$ measurement over the frequency range of $2$ to $10$ GHz. Figure~\ref{fig:s21} shows the results, where the $S_{21}$ measurement of the undeformed \ac{RIS} indicates its expected reflective behavior across nearly the entire frequency range. Additionally, the $S_{21}$ measurement for the deformed \ac{RIS} is also illustrated in Figure~\ref{fig:s21}. For the deformed \ac{RIS} measurements, a structural deformation was introduced at the center of the \ac{RIS}, by using a dielectric material to create a localized surface displacement. This resulted in a surface curvature of approximately $1$ cm (e.g. $0.1\lambda$ for $3$ GHz), simulating structural deformations that may arise from environmental factors or installation processes. As observed in Figure~\ref{fig:s21}, the deformation leads to a measurable deviation from the obtained with the undeformed \ac{RIS}. More precisely, one can verify a deviation with a \ac{MSE} of $0.0024$ or, correspondingly, of $-26.27$ dB between the deformed and undeformed measurements. Therefore, these results highlight the sensitivity of the \ac{RIS} performance to \acp{HWI}. In other words, the measured $S_{21}$ magnitude revealed noticeable changes across the frequency spectrum when there is even a slight deformation on the \ac{RIS} surface. This confirms that even minor material variations can cause significant perturbations in the \ac{RIS} response, which, if not accounted for, may degrade the system performance.

\section{Numerical Results}\label{sec:numresult}
As stated in Section~\ref{sec:grad}, we leverage numerical methods for computing the gradients that are employed in the gradient descent optimization. In the following, we discuss how the optimization is carried out. Furthermore, the compensated \ac{RIS} configuration provided by the proposed gradient descent optimizer is evaluated using computational simulations.

However, let us first specify all relevant system model parameters, since they are used throughout this section: (i) the $L_a = 101$ path delays $\tau^l_a \sim \mathcal{U} \left[\tau^1_a\text{,}2\tau^1_a\right]\text{, } \forall l > 1$, have their first path, $l = 1$, as the strongest \ac{LOS} path (same apply for the $L_b = 51$ path delays $\tau^\ell_b$); (ii) all $L_d = 100$ path delays obey $\tau^l_d \sim \mathcal{U} \left[\tau_d\text{,}2\tau_d\right]$, where $\tau_d$ is the time, in seconds, that the signal takes to propagate form the \ac{AP} to the \ac{UE} (\ac{LOS} path); (iii) the azimuth and elevation angles of arrival/departure considered for the \ac{RIS} can vary randomly around the \ac{LOS} path initial angle, with $\boldsymbol{\phi}_{a,b} \sim \mathcal{U} \left[-40^{\circ}\text{,}40^{\circ}\right]$ and $\boldsymbol{\varphi}_{a,b} \sim \mathcal{U} \left[-10^{\circ}\text{,}10^{\circ}\right]$; (iv) also let $f_c = 3$ GHz, $B = 10.5$ MHz and $N_0=-164$ dBm; and, finally, (v) we assume for each \ac{RIS} reflector that $d_\text{H} = d_\text{V} = 0.25\lambda$ meters. We moreover refer the interested readers to \cite{pedro:25} for further details about the channel model specifications.

\subsection{Gradient Descent Optimizer}
In this work, we employ the so-called automatic differentiation in order to compute the gradients expressed in \eqref{eq:grad}. More specifically, we use the computational tools provided by the TensorFlow platform \cite{tensorflow:16}. With the gradients computed by the automatic differentiation, then the phase-shift compensation, $\boldsymbol{\bar{\theta}}$, is updated in the following manner:
\begin{equation}\label{eq:upd}
\boldsymbol{\bar{\theta}}_{k} = \boldsymbol{\bar{\theta}}_{k-1} - \gamma\nabla_{\boldsymbol{\bar{\theta}}},
\end{equation}
where $k = [1\text{,}2\text{, }\dots\text{,}K]^\text{T}$ and $\gamma$ roughly controls by how much, or how fast, is the descent to local minima of \eqref{eq:grad}; it is usually referred to as the learning rate. This update operation is repeated until the stopping criteria is met or if the maximum number of iterations, $K$, is reached.

The gradients update shown in \eqref{eq:upd} is one of the simplest form of computing the gradient descent, with more sophisticated alternatives such as the ADAM optimizer being also available. For readers interested in an comprehensive discussion of different optimizations methods, see \cite{zapp:19} and the references therein. We nevertheless briefly show that the \ac{RIS} hardware imperfections can be compensated more efficiently with the simple update given by \eqref{eq:upd}. 

Let us first introduce the \ac{STM} \cite{pedro:25,bjorn:22} method for configuring the \ac{RIS}:
\begin{equation}\label{eq:stmop}
    \boldsymbol{\omega}_{\theta}^{(m^\ast)} = \underset{m \in \left\{0,1,\dots,M-1\right\}}{\arg \max}{\|h_d\left[m\right] + \mathbf{V}_m^\text{T} \boldsymbol{\omega}_{\theta}^{(m)} \|_2^2}\text{,}
\end{equation}
where $\mathbf{V}_m$ is the $m$-th column of $\mathbf{V}$ and also
\begin{equation}\label{eq:stmphase}
    \boldsymbol{\omega}_{\theta_n}^{(m)} = e^{\jmath \left(\arg{\left\{h_d\left[m\right]\right\}} - \arg{\left\{V_{m,n}\right\}}\right)} \ , \ \forall n \in \{0\text{,}1\text{,}\ldots\text{,}N-1\}\text{,}
\end{equation}
which represents the alignment of phase-shifts for all $N$ \ac{RIS} elements, such that the direct channel, $\mathbf{h}_d$, combines in-phase with the composite channel, $\mathbf{V}$, for the $m$-th time sample. Therefore, in this work we assume that the \ac{STM} is the ideal phase-shift \ac{RIS} configuration as intended by the \ac{AP}. However, from Section~\ref{sec:hdwImperf}, we know that this configuration is not going to be ideally represented by the \ac{RIS} reflectors.
 
\begin{figure}[h!]
	\centering
	\includegraphics[width=0.975\columnwidth,keepaspectratio]{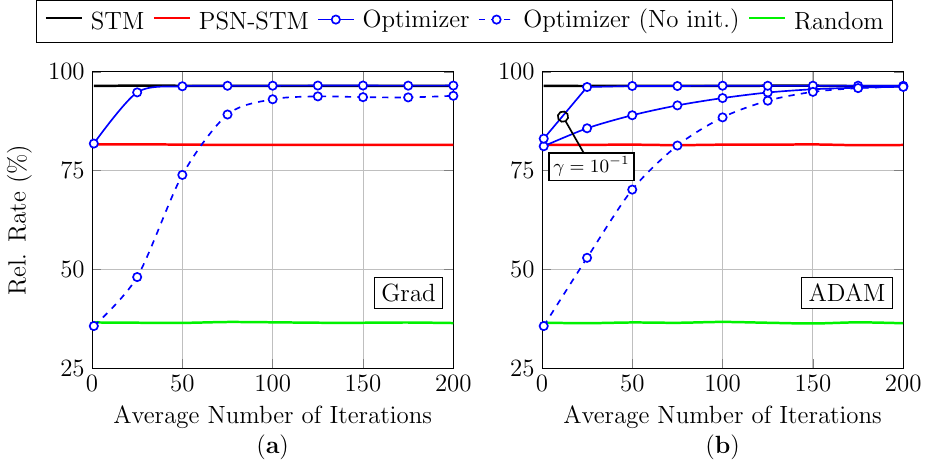}
	\caption{Relative rate performance of the gradient descent (`Grad') and ADAM optimizers, in comparison to the ideal/uncompensated \ac{STM} configurations for the \ac{RIS}. Each simulation point employed up to $5\times10^3$ Monte Carlo runs.}\label{fig:iterPlot}
\end{figure}

With that in mind, observe in Figure~\ref{fig:iterPlot} that the relative rate\footnote{The relative rate is the ratio between the best achievable rate obtained with perfectly coherent combination, and the actual achievable rate \cite{pedro:25}.} obtained from the optimization of a imperfect \ac{STM} configuration, is traced for the average number of gradient update iterations (with $\gamma = 10^{-2}$). The ideal \ac{STM} configuration performance and the \ac{PSN} effect on its performance is also illustrated in Figure~\ref{fig:iterPlot}, alongside the results obtained with a random compensator\footnote{We assume a random compensator phase $\boldsymbol{\bar{\theta}} \sim \mathcal{U} [-\pi\text{,}\pi]$, instead of the one computed by the optimizer, as a lower bound for performance.}. Consequently, note that both the gradient descent and ADAM optimizers improve their performances as the number of iterations increases, whilst the performances for the \ac{STM} and random compensator are fixed, as expected. The objective of the optimizers is to compensate for hardware imperfections (see Section~\ref{sec:grad}), so that it can be seen in Figure~\ref{fig:iterPlot} that after $\sim 25$ iterations the optimizers are able to reach the ideal \ac{STM} configuration performance. In fact, the ADAM optimizer can be even employed in the \ac{RIS} configuration by itself (Figure~\ref{fig:iterPlot} (b)), that is, with no initialization from the \ac{STM} configuration ($\boldsymbol{\omega}^\prime_{\theta} = \boldsymbol{\hat{\omega}}_{\theta}$ since only the \ac{PSN} is present). In conclusion, although the updates of \eqref{eq:upd} do not offer the most robust optimizer performance, yet for the task at hand, that is, hardware imperfections compensation, it is equivalent to the ADAM optimizer. Since the complexity of the ADAM is greater \cite{zapp:19} for each iteration, thus the simple gradient descent update of \eqref{eq:upd} is used for the remainder of this work.

\subsection{Performance Evaluation}
In this subsection we evaluate the performance of the gradient descent optimizer, utilizing the gradients update operation given by \eqref{eq:upd}. 

\begin{figure}[h!]
	\centering
	\includegraphics[width=0.575\columnwidth,keepaspectratio]{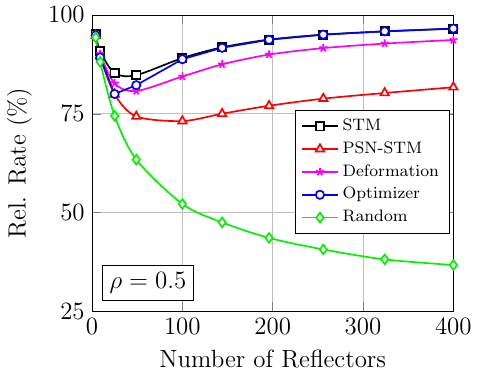}
	\caption{Relative rate performance for the optimizer \eqref{eq:upd}, considering the \ac{STM} affected by the \ac{PSN} \eqref{eq:errmodel} and also by the \ac{RIS} surface deformations \eqref{eq:defmodel}, with the ideal \ac{STM} \eqref{eq:stmop} as a upper bound reference and the random configuration as the lower bound. Each simulation point employed $10^3$ Monte Carlo runs.}\label{fig:piPlot}
\end{figure}

Figure~\ref{fig:piPlot} shows the relative rate performance with the optimizer, in comparison to the ideal performance obtained with the \ac{STM} configuration of a perfect \ac{RIS}. Moreover, Figure~\ref{fig:piPlot} illustrates the \ac{STM} configuration performance considering the hardware imperfections discussed in Section~\ref{sec:hdwImperf}, that is, the \ac{PSN} of \eqref{eq:errmodel} and the \ac{RIS} surface deformations of \eqref{eq:defmodel}. It is worth noting that the optimizer compensates the \ac{RIS} configuration under the combined effect of the aforementioned hardware imperfections. Note also that the random configuration is presented in Figure~\ref{fig:piPlot}, being a lower bound to the relative rate performance. Finally, we assume $\rho = 0.5$ in \eqref{eq:errmodel}, $h_{\text{max}} = 0.1\lambda$ \cite{yang:23} and $k = \pi$ in \eqref{eq:defmodel}. The \ac{LOS} path initial angle is such that $\boldsymbol{\varphi}_{a,b} = 0^{\circ}$; we also employ $K = 700$ subcarriers and a range of \ac{RIS} sizes ($N$).     

\begin{figure}[ht!]
	\centering
	\includegraphics[width=0.999\columnwidth,keepaspectratio]{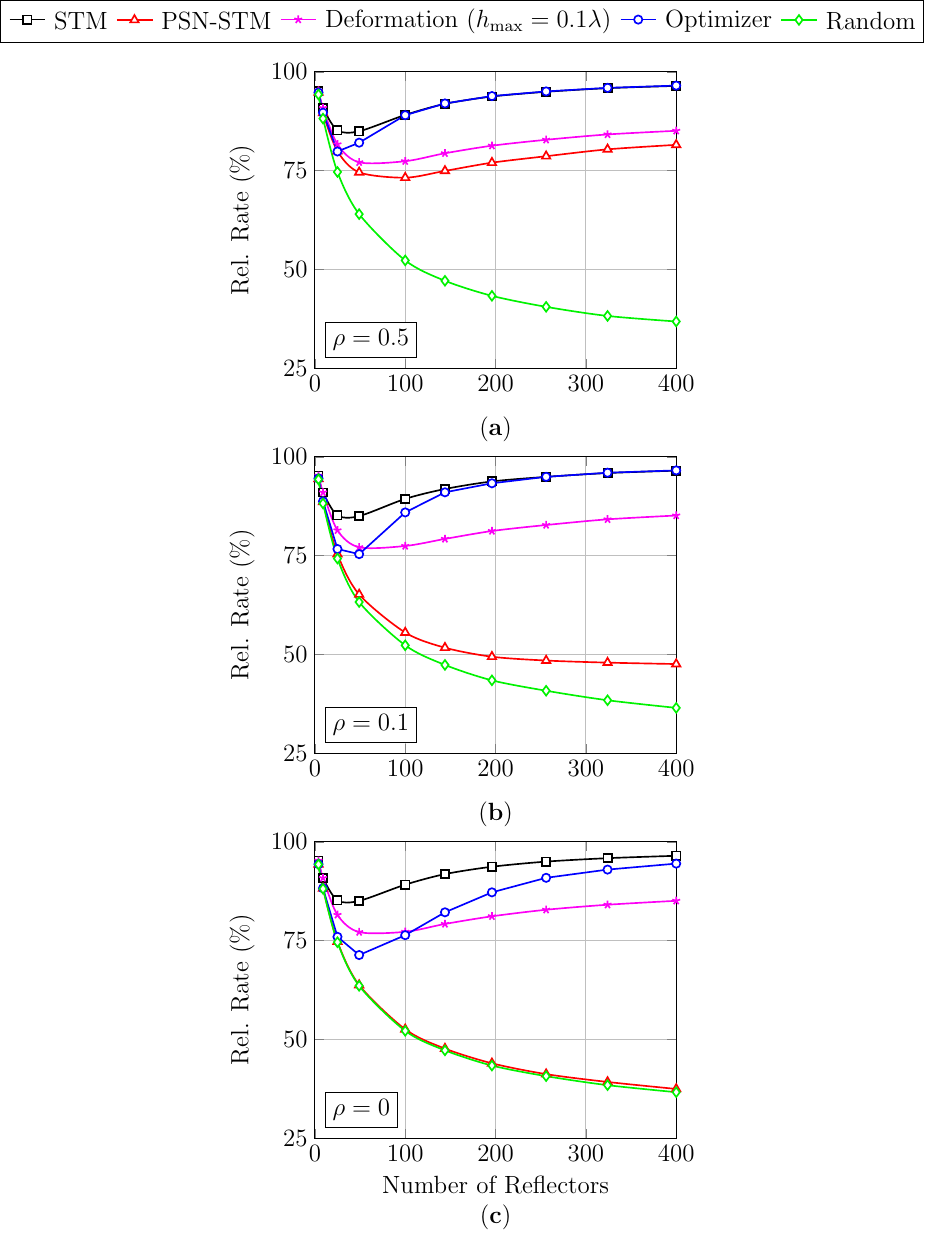}
	\caption{Relative rate performance for the optimizer \eqref{eq:upd}, now compensating for the combined hardware imperfections considering \ac{RIS} surfaces deformations with $k = 2\pi$ in \eqref{eq:defmodel}. Also, $\rho = 0.5$ (a), $\rho = 0.1$ (b) and $\rho = 0$ (c) in \eqref{eq:errmodel} are evaluated. Each simulation point employed up to $5\times 10^3$ Monte Carlo runs.}\label{fig:0_1Plot}
\end{figure}

Therefore, observe in Figure~\ref{fig:piPlot} that the optimizer is able to totally compensate the hardware imperfections of \acp{RIS} with $100$ reflectors or more. Despite the considerable drop in performance due to the \ac{PSN}, the \ac{RIS} deformations on the other hand are not as impactful. Consequently, in Figure~\ref{fig:0_1Plot} we evaluate the relative rate performance for $k = 2\pi$ in \eqref{eq:defmodel}, also with $\rho = 0.5$ (Figure~\ref{fig:0_1Plot} (a)), $\rho = 0.1$ (Figure~\ref{fig:0_1Plot} (b)) and $\rho = 0$ (Figure~\ref{fig:0_1Plot} (c)), while other parameters are kept the same. It is shown in \cite{yang:23} that increasing $k$ can lead to beamforming shapes that are detrimental to the general \ac{RIS} performance. 

Therefore, verify in Figure~\ref{fig:0_1Plot} (a) that increasing the number of deformation peaks, $k$, also worsens the relative rate performance. We conjecture that this happens because higher values of $k$ translates to more phase-shift discrepancy between adjacent groups of \ac{RIS} reflectors. This in turn will lead to more phase misalignment at the \ac{UE}, consequently reducing the achievable rate. Note also in Figures~\ref{fig:0_1Plot} (b)-(c) that the increase in the \ac{PSN} intensity can lead to prohibitive performances by the \ac{STM} configuration for the \ac{RIS}. Nevertheless, the optimizer is still able to compensate satisfactorily for the combined hardware imperfections. See in Figure~\ref{fig:0_1Plot} (c) that even with the fully uncorrelated \ac{PSN}, the optimizer can reach the ideal \ac{STM} performance for a \ac{RIS} with $N=400$ reflectors. 

\begin{figure}[ht!]
	\centering
	\includegraphics[width=0.97\columnwidth,keepaspectratio]{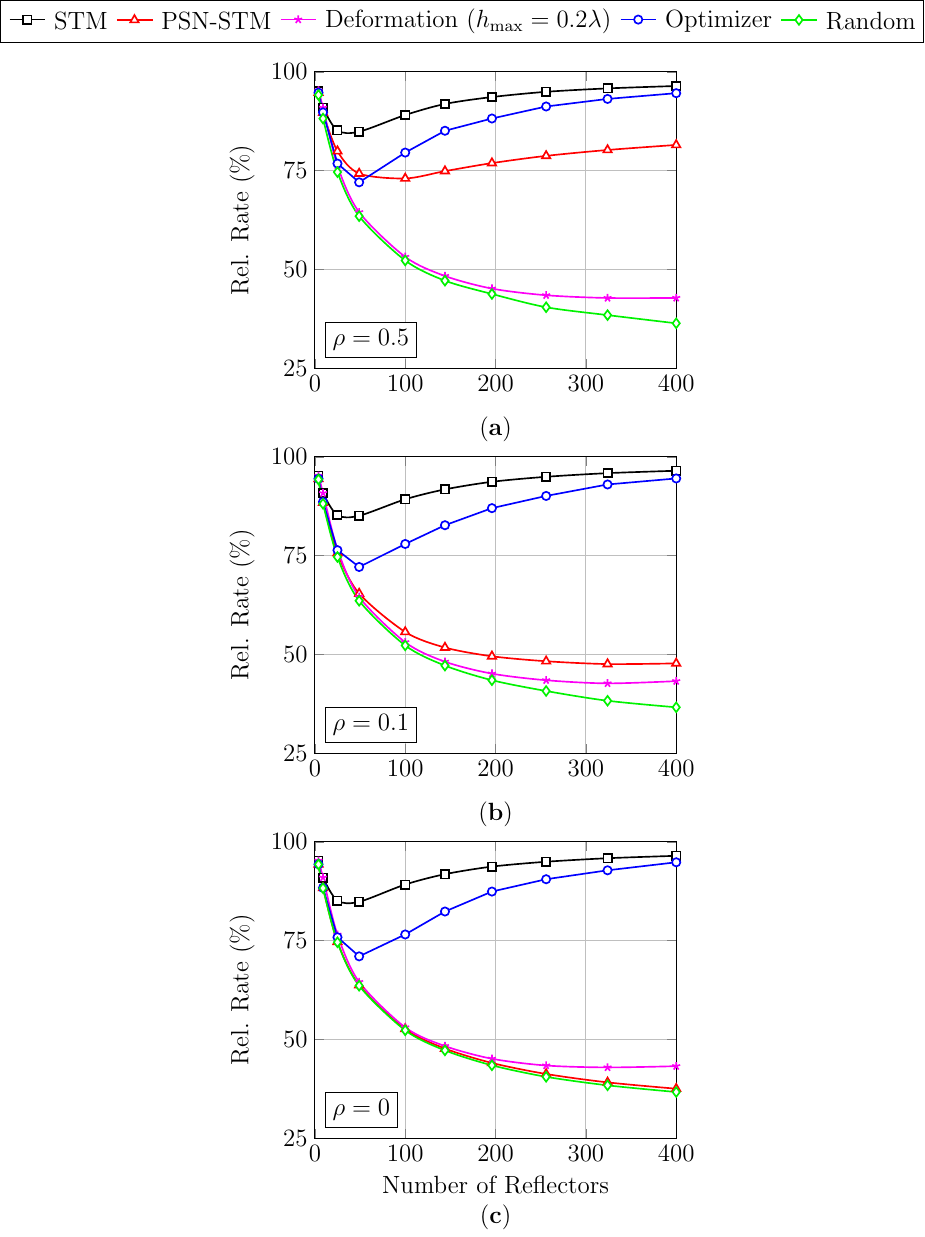}
	\caption{Relative rate performance for the optimizer \eqref{eq:upd}, now considering \ac{RIS} surfaces deformations with peaks magnitudes defined by $h_{\text{max}} = 0.2\lambda$ in \eqref{eq:defmodel}. Each simulation point employed up to $5\times 10^3$ Monte Carlo runs.}\label{fig:0_2Plot}
\end{figure}

Additionally, Figure~\ref{fig:0_2Plot} shows the relative rate performance changes caused by only increasing the magnitude of deformation peaks to $h_{\text{max}} = 0.2\lambda$. As reported in \cite{yang:23}, for $k=2\pi$, values of $h_{\text{max}} > 0.1\lambda$ can lead to significant loss of beamforming performance. This is also true for the relative rate performance, specially when the combined hardware imperfections are evaluated, as in Figures~\ref{fig:0_2Plot} (a)-(c). Yet note that the optimizer is able to compensate these severe hardware imperfections, reaching the ideal \ac{STM} performance for $N=400$ reflectors.             
%

\section{Conclusions}\label{sec:conclusion}

\commenttxt{This work presents a lightweight, numerically driven gradient descent optimization framework for mitigating \acp{HWI} in \ac{RIS} assisted wideband communication systems. By leveraging automatic differentiation, the proposed approach removes the need for analytical gradient derivations or prior statistical modeling of hardware imperfections, providing a flexible, scalable, and implementation-friendly alternative to existing optimization methods. Comprehensive numerical and experimental results demonstrated that the proposed method effectively compensates for key \acp{HWI}, including \ac{PSN} and \ac{RIS} surface deformations, recovering the system’s achievable rate to levels comparable to ideal, imperfection-free conditions. The optimizer exhibited strong adaptability across varying levels of \acp{HWI} severity and \ac{RIS} sizes, confirming its suitability for real-time deployment in dynamic wireless environments.} 

\commenttxt{Beyond performance recovery, the findings of this study highlight a paradigm shift in how signal optimization can be addressed in next-generation communication architectures. By bridging electromagnetic-level imperfections with system-level adaptability, the proposed framework offers a practical pathway toward robust and self-calibrating \ac{RIS} operation. Future research can build upon this foundation by developing customized automatic differentiation frameworks tailored specifically for \acp{HWI} compensation and by conducting broader experimental validations. Such advancements will contribute to more precise hardware characterization and further accelerate the practical adoption of \ac{RIS} technologies in \ac{6G} networks.}


\bibliography{references}

@ARTICLE{alexandropoulos:22,
  author={Alexandropoulos, George C. and Stylianopoulos, Kyriakos and Huang, Chongwen and Yuen, Chau and Bennis, Mehdi and Debbah, Mérouane},
  journal={Proceedings of the IEEE}, 
  title={Pervasive Machine Learning for Smart Radio Environments Enabled by Reconfigurable Intelligent Surfaces}, 
  year={2022},
  volume={110},
  number={9},
  pages={1494-1525},
  doi={10.1109/JPROC.2022.3174030}
}

@misc{astr:24,
      title={{RIS} in Cellular Networks -- Challenges and Issues}, 
      author={Magnus Åström and Philipp Gentner and Omer Haliloglu and Behrooz Makki and Ola Tageman},
      year={2024},
      eprint={2404.04753},
      archivePrefix={arXiv},
      primaryClass={cs.NI},
      url={https://arxiv.org/abs/2404.04753}, 
}

@ARTICLE{bjorn:22,
    author={Björnson, Emil and Wymeersch, Henk and Matthiesen, Bho and Popovski, Petar and Sanguinetti, Luca and de Carvalho, Elisabeth},
    journal={IEEE Signal Process. Mag.}, 
    title={{R}econfigurable {I}ntelligent {S}urfaces: {A} signal processing perspective with wireless applications}, 
    year={2022},
    volume={39},
    number={2},
    pages={135-158},
    doi={10.1109/MSP.2021.3130549}
}

@ARTICLE{bjorn:13,
    author={Bjornson, Emil and Zetterberg, Per and Bengtsson, Mats and Ottersten, Bjorn},
    journal={IEEE Commun. Lett.}, 
    title={Capacity Limits and Multiplexing Gains of {MIMO} Channels with Transceiver Impairments}, 
    year={2013},
    volume={17},
    number={1},
    pages={91-94},
    doi={10.1109/LCOMM.2012.112012.122003}
}

@ARTICLE{feng:21,
    author={Feng, Biqian and Gao, Junyuan and Wu, Yongpeng and Zhang, Wenjun and Xia, Xiang-Gen and Xiao, Chengshan},
    journal={IEEE Wirel. Commun.}, 
    title={Optimization Techniques in Reconfigurable Intelligent Surface Aided Networks}, 
    year={2021},
    volume={28},
    number={6},
    pages={87-93},
    doi={10.1109/MWC.001.2100196}
}

@INPROCEEDINGS{huang:24,
  author={Huang, Sin-Yu and Lin, Jia-You and Wang, Chih-Yu and Hwang, Ren-Hung},
  booktitle={2024 IEEE 99th Vehicular Technology Conference (VTC2024-Spring)}, 
  title={{MSE} Minimization for {RIS}-Assisted Wireless Networks with Phase Error and Phase-Dependent Amplitude Response}, 
  year={2024},
  volume={},
  number={},
  pages={1-6},
  doi={10.1109/VTC2024-Spring62846.2024.10683024}
}

@article{jorge:25,
title   = {Electric dipole-magnetic quadrupole resonant coupling for broadband operating metasurfaces},
author = {Ribeiro, Jéssica Abranches Pinto and Hernandez-Sarria, Jhon James and Pereira, Luiz Augusto Melo and Sodré, Arismar Cerqueira and Oliveira Junior, Osvaldo Novais de and Mendes, Luciano Leonel and Mejía-Salazar, Jorge Ricardo},
year = {2025},
doi = {10.1109/TAP.2024.3515260},
journal   = {IEEE Transactions on Antennas and Propagation}
}

@inproceedings{liX:25,
  title={Potential Standardization Work for Reconfigurable Intelligent Surface White Paper},
  author={Li, NX and Yuan, YF and Liu, QY and Zhao, YJ and Jin, S and others},
  booktitle={FuTURE Forum, Nanjing, China},
  year={2025}
}

@ARTICLE{liaskos:22,
    author={Liaskos, Christos and Mamatas, Lefteris and Pourdamghani, Arash and Tsioliaridou, Ageliki and Ioannidis, Sotiris and Pitsillides, Andreas and Schmid, Stefan and Akyildiz, Ian F.},
    journal={Proc. IEEE}, 
    title={{S}oftware-{D}efined {R}econfigurable {I}ntelligent {S}urfaces: {F}rom Theory to End-to-End Implementation}, 
    year={2022},
    volume={110},
    number={9},
    pages={1466-1493},
    doi={10.1109/JPROC.2022.3169917}
}

@ARTICLE{pedro:25,
  author={de Souza, Pedro H. C. and Khazaee, Masoud and Leonel Mendes, Luciano},
  journal={IEEE Transactions on Communications}, 
  title={Resource-Efficient Configuration of {RIS}-Aided Communication Systems Under Discrete Phase-Shifts and User Mobility}, 
  year={2025},
  volume={73},
  number={1},
  pages={145-157},
  doi={10.1109/TCOMM.2024.3432690}
}

@misc{tensorflow:16,
      title={TensorFlow: A system for large-scale machine learning}, 
      author={Martín Abadi and Paul Barham and Jianmin Chen and Zhifeng Chen and Andy Davis and Jeffrey Dean and Matthieu Devin and Sanjay Ghemawat and Geoffrey Irving and Michael Isard and Manjunath Kudlur and Josh Levenberg and Rajat Monga and Sherry Moore and Derek G. Murray and Benoit Steiner and Paul Tucker and Vijay Vasudevan and Pete Warden and Martin Wicke and Yuan Yu and Xiaoqiang Zheng},
      year={2016},
      eprint={1605.08695},
      archivePrefix={arXiv},
      primaryClass={cs.DC},
      url={https://arxiv.org/abs/1605.08695}, 
}

@ARTICLE{xian:25,
  author={Zhao, Xianming and Jian, Mengnan and Chen, Yijian and Zhao, Yajun and Mu, Lin},
  journal={Intelligent and Converged Networks}, 
  title={Reconfigurable Intelligent Surfaces for {6G}: Engineering Challenges and the Road Ahead}, 
  year={2025},
  volume={6},
  number={1},
  pages={53-81},
  doi={10.23919/ICN.2025.0004}
}

@ARTICLE{wang:24,
  author={Wang, Ruiqi and Yang, Yiming and Makki, Behrooz and Shamim, Atif},
  journal={IEEE Transactions on Antennas and Propagation}, 
  title={A Wideband Reconfigurable Intelligent Surface for {5G} Millimeter-Wave Applications}, 
  year={2024},
  volume={72},
  number={3},
  pages={2399-2410},
  doi={10.1109/TAP.2024.3352828}
}

@ARTICLE{yang:20,
    author={Yang, Yifei and Zheng, Beixiong and Zhang, Shuowen and Zhang, Rui},
    journal={IEEE Trans. Commun.}, 
    title={Intelligent Reflecting Surface Meets {OFDM}: Protocol Design and Rate Maximization}, 
    year={2020},
    volume={68},
    number={7},
    pages={4522-4535},
    doi={10.1109/TCOMM.2020.2981458}
}

@ARTICLE{yang:23,
  author={Yang, Jun and Chen, Yijian and Cui, Yijun and Wu, Qingqing and Dou, Jianwu and Wang, Yuxin},
  journal={IEEE Transactions on Communications}, 
  title={How Practical Phase-Shift Errors Affect Beamforming of Reconfigurable Intelligent Surface?}, 
  year={2023},
  volume={71},
  number={10},
  pages={6130-6145},
  doi={10.1109/TCOMM.2023.3293859}
}

@ARTICLE{yue:24,
  author={Yue, Chengfang and Tang, Hui and Chai, Li},
  journal={IEEE Transactions on Vehicular Technology}, 
  title={Fast Beamforming for {IRS} Assisted Multi-User Communication Systems by Lightweight Unsupervised Learning}, 
  year={2024},
  volume={73},
  number={11},
  pages={17180-17191},
  doi={10.1109/TVT.2024.3427001}
}

@ARTICLE{zapp:19,
    author={Zappone, Alessio and Di Renzo, Marco and Debbah, Mérouane},
    journal={IEEE Transactions on Communications}, 
    title={Wireless Networks Design in the Era of Deep Learning: Model-Based, {AI}-Based, or Both?}, 
    year={2019},
    volume={67},
    number={10},
    pages={7331-7376},
    doi={10.1109/TCOMM.2019.2924010}
}

@article{pan2023phased,
  title={{Phased Array Antenna Calibration Method Experimental Validation and Comparison}},
  author={Pan, Chong and Ba, Xinran and Tang, Yuanhua and Zhang, Fengchun and Zhang, Yusheng and Wang, Zhengpeng and Fan, Wei},
  journal={Electronics},
  volume={12},
  number={3},
  pages={489},
  year={2023},
  publisher={MDPI}
}

@book{pozar2021microwave,
  title={Microwave Engineering: Theory and Techniques},
  author={Pozar, David M},
  year={2021},
  publisher={John Wiley \& Sons},
  address   = {Hoboken, New Jersey}
}

@inproceedings{zhang2024phases,
  title={{Phases Calibration of RIS Using Backpropagation Algorithm}},
  author={Zhang, Wei and Zhou, Bin and Zhang, Tianyi and Jiang, Yi and Bu, Zhiyong},
  booktitle={2024 IEEE/CIC International Conference on Communications in China (ICCC)},
  pages={445--449},
  year={2024},
  organization={IEEE}
}
\bibliographystyle{IEEEtran}

\begin{IEEEbiography}[{\includegraphics[width=1in,height=1.25in,clip,keepaspectratio]{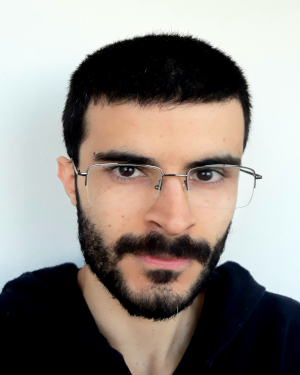}}]{Pedro H. C. de Souza} was born in Santa Rita do Sapuca\'i, Minas Gerais, MG, Brazil in 1992. He received the B.S., M.S. and the Doctor degrees in telecommunications engineering from the National Institute of Telecommunications - INATEL, Santa Rita do Sapuca\'i, in 2015, 2017 and 2022, respectively; is currently working as a postdoctoral researcher in telecommunications engineering at INATEL, with the support of FAPESP (\textit{Fundação de Amparo à Pesquisa do Estado de São Paulo}). During the year of 2014 he was a Hardware Tester with the INATEL Competence Center - ICC. His main interests are: digital communication systems, mobile telecommunications systems, 6G, reconfigurable intelligent surfaces, convex optimization for telecommunication systems, compressive sensing/learning, cognitive radio.
\end{IEEEbiography}

\begin{IEEEbiography}[{\includegraphics[width=1in,height=1.25in,clip,keepaspectratio]{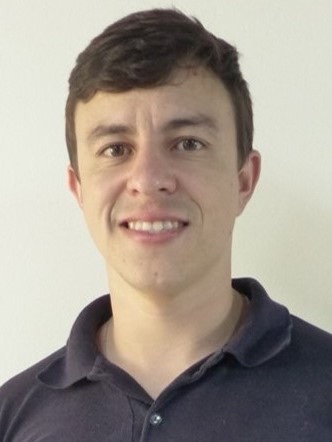}}]{Luiz A. M. Pereira} received the B.Sc., M.Sc., and Ph.D. degrees in Telecommunications from the National Institute of Telecommunications (Inatel), Brazil, in 2017, 2020, and 2023, respectively. He has acted as a researcher at the Wireless and Optical Convergent Access (WOCA) and Reference Center in Radiocommunications (CRR) laboratories, working on Brasil 6G and xGMobile Projects. Since 2024, he is a professor at Inatel.
\end{IEEEbiography}

\begin{IEEEbiography}[{\includegraphics[width=1in,height=1.25in,clip,keepaspectratio]{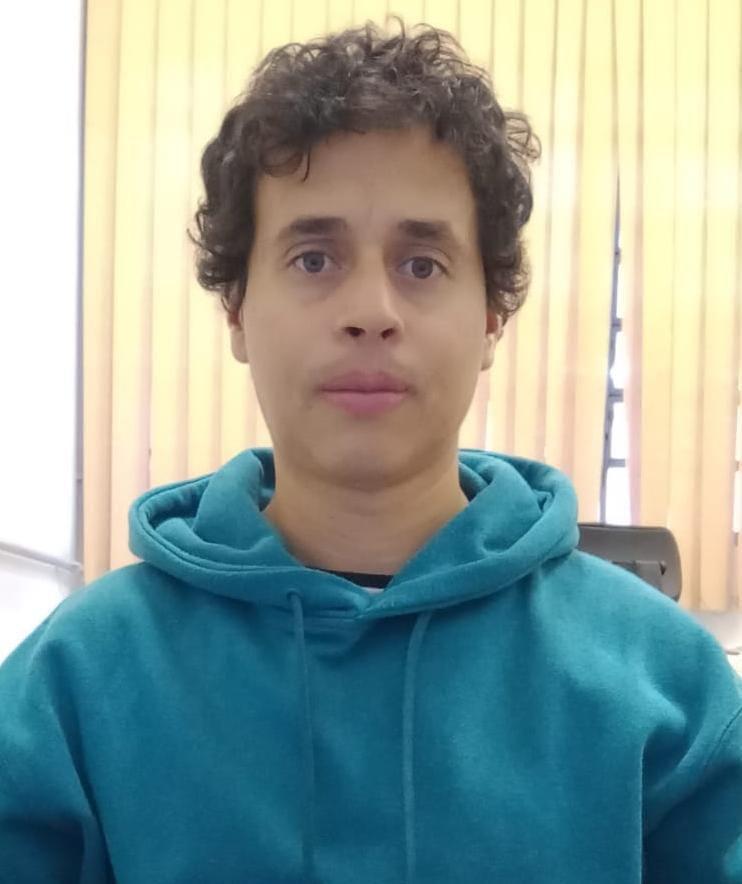}}]{Faustino Reyes Gómez} received his B.Sc., M.Sc., and Ph.D. in Physics from Universidad del Valle in Cali, Colombia. He is currently a postdoctoral researcher at the National Telecommunications Institute (Inatel). His main research areas include the design and simulation of nanophotonic systems for biosensing applications, as well as the design, simulation, and development of metasurfaces for use in 5G and 6G telecommunications. He also has teaching experience, having served as a teaching assistant and full-time professor in various physics courses.
\end{IEEEbiography}

\begin{IEEEbiography}[{\includegraphics[width=1in,height=1.25in,clip,keepaspectratio]{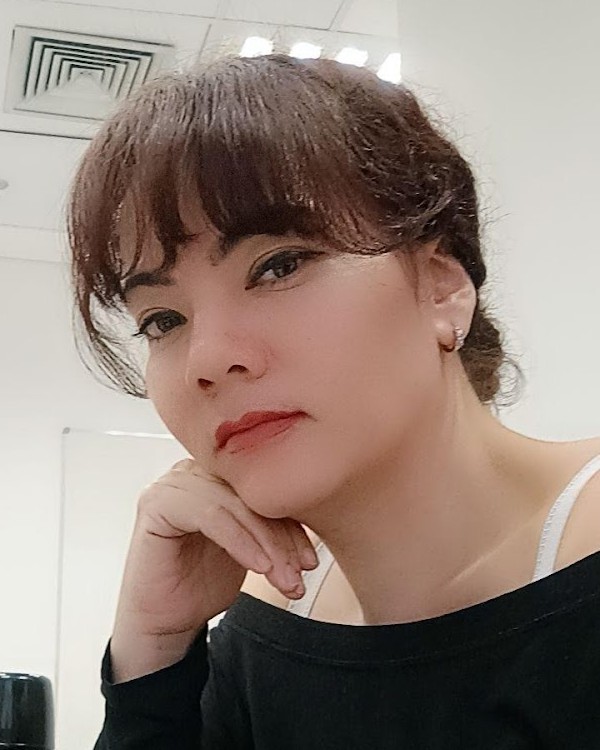}}]{Elsa Maria Materón} holds a degree in Chemistry from Universidad del Valle (Colombia), where she excelled in her work on condensed matter physics. She earned both her Master’s and Ph.D. in Biotechnology from São Paulo State University (UNESP). During her doctoral studies, she completed research internships at the Institute for Nanotechnology at the University of Waterloo (Canada), focusing on the SELEX process for aptamer synthesis and the development of enzymatic biosensors for studying chemotherapeutic drugs. 

Additionally, she was a visiting student at the University of Pennsylvania (USA), where she worked on graphene-based field-effect transistors and the synthesis of graphene via CVD. She completed postdoctoral fellowships at the Federal University of São Carlos (UFSCar), the University of Pennsylvania (Department of Physics and Astronomy), the Institute of Physics of São Carlos (IFSC-USP), the Institute of Chemistry of São Carlos (IQSC-USP), and the Graphene and Nanomaterials Research Center (MackGraphe). She is a researcher at the National Institute of Telecommunications (Inatel) and a collaborator at IFSC. Her research focuses on developing electrochemical and optical sensors and biosensors for clinical diagnostics, microfluidic devices, microfabrication, and antenna and metasurface fabrication. She possesses extensive experience in nanomaterial synthesis, biogenic nanoparticles, Langmuir films, and the physicochemical characterization of surfaces.
\end{IEEEbiography}

\begin{IEEEbiography}[{\includegraphics[width=1in,height=1.25in,clip,keepaspectratio]{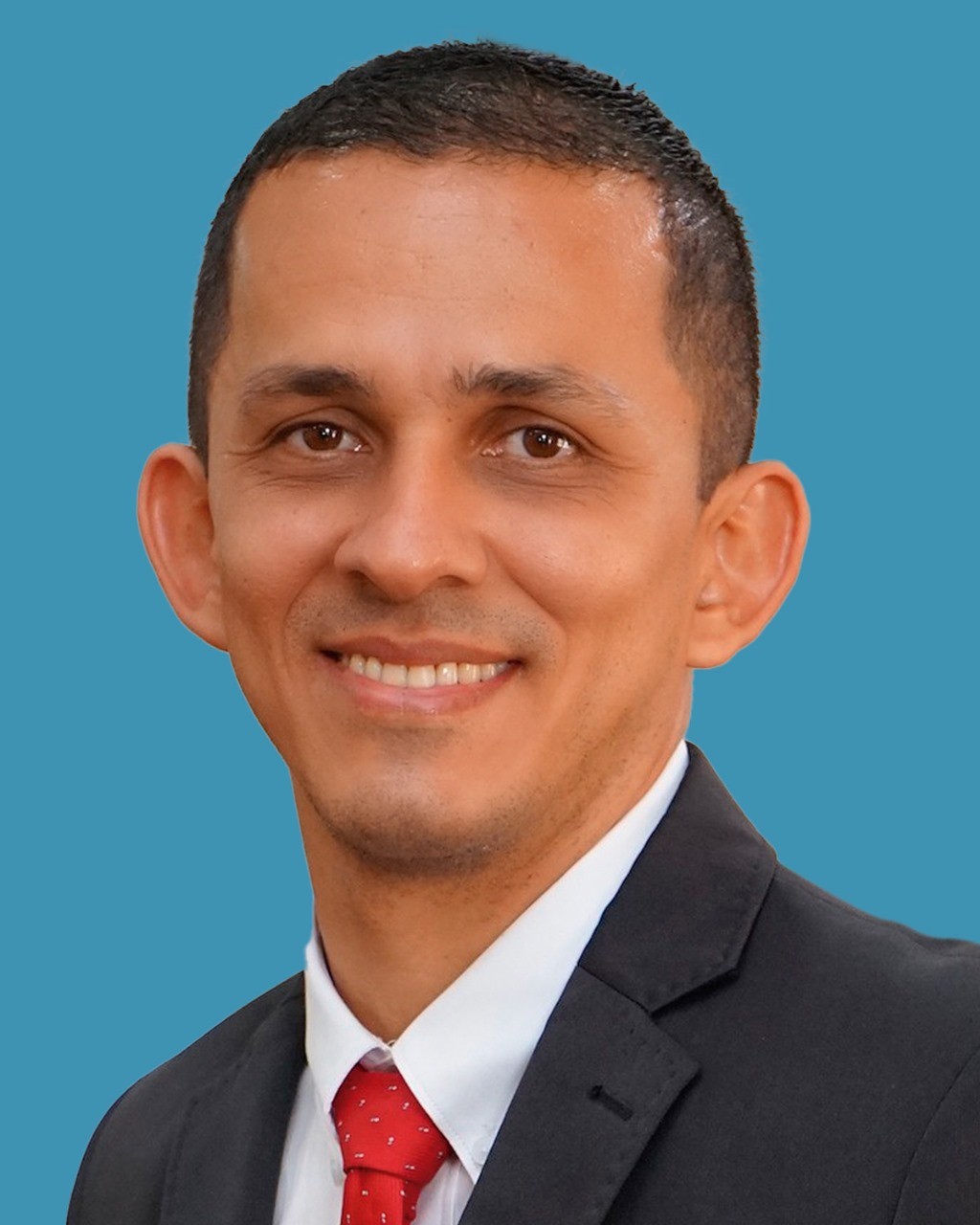}}]{Jorge Ricardo Mej\'ia-Salazar} obtained his B.Sc., M.Sc., and Ph.D. degrees in Physics from Universidad del Valle in Cali, Colombia, in 2008, 2009, and 2014, respectively. After completing his Ph.D., he pursued postdoctoral research at the Institute of Physics, Federal University of Alagoas, Macei\'o, Brazil (2014–2016), and later at the Sao Carlos Institute of Physics, University of Sao Paulo, Brazil (2016–2018). Since 2018, Jorge has been a Professor at the National Institute of Telecommunications (Inatel) in Santa Rita do Sapuca\'i, Minas Gerais, Brazil. His research interests include applied electromagnetism, reconfigurable intelligent surfaces (RIS), antennas, nanoantennas, and the electromagnetic principles governing chiral, magneto-chiral, magneto-optic, and magnetoplasmonic effects in nanostructures.
\end{IEEEbiography}

\begin{IEEEbiography}[{\includegraphics[width=1in,height=1.25in,clip,keepaspectratio]{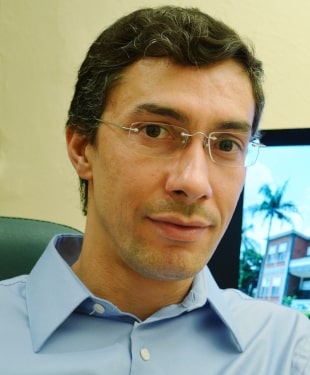}}]{Luciano Leonel Mendes} received the B.Sc. and M.Sc. degrees from Inatel, Brazil, in 2001 and 2003, respectively, and the Doctor degree from Unicamp, Brazil, in 2007, all in electrical engineering. Since 2001, he has been a Professor with Inatel, where he has acted as the Technical Manager of the Hardware Development Laboratory from 2006 to 2012. From 2013 to 2015, he was a Visiting Researcher with the Technical University of Dresden in the Vodafone Chair Mobile Communications Systems, where he has developed his postdoctoral. In 2017, he was elected Research Coordinator of the 5G Brazil Project, an association involving industries, telecom operators, and academia which aims for funding and build an ecosystem toward 5G in Brazil. He is also the technical coordinator of the Brazil 6G Project.
\end{IEEEbiography}




\EOD

\end{document}